\documentclass[prl,twocolumn,floatfix,superscriptaddress]{revtex4}
\usepackage{graphicx}
\usepackage[light,first,bottomafter]{draftcopy}

\newcommand{\be}{\begin{equation}}
\newcommand{\ee}{\end{equation}}
\newcommand{\ba}{\begin{eqnarray}}
\newcommand{\ea}{\end{eqnarray}}
\newcommand{\nn}{\nonumber \\}

\begin{document}

 \title{Probing High Frequency Noise with Macroscopic Resonant Tunneling}
\author{T. Lanting}
\author{M.~H.~S.~Amin}
\address{D-Wave Systems Inc., 100-4401 Still Creek Drive, Burnaby, B.C., V5C 6G9, Canada}
\author{M.~W.~Johnson}
\address{D-Wave Systems Inc., 100-4401 Still Creek Drive, Burnaby, B.C., V5C 6G9, Canada}
\author{F.~Altomare}
\address{D-Wave Systems Inc., 100-4401 Still Creek Drive, Burnaby, B.C., V5C 6G9, Canada}
\author{A.~J.~Berkley}
\address{D-Wave Systems Inc., 100-4401 Still Creek Drive, Burnaby, B.C., V5C 6G9, Canada}
\author{S.~Gildert}
\address{D-Wave Systems Inc., 100-4401 Still Creek Drive, Burnaby, B.C., V5C 6G9, Canada}
\author{R.~Harris}
\address{D-Wave Systems Inc., 100-4401 Still Creek Drive, Burnaby, B.C., V5C 6G9, Canada}
\author{J.~Johansson}
\address{Department of Natural Sciences, University of Agder, 4604 Kristiansand, Norway}
\author{P.~Bunyk}
\address{D-Wave Systems Inc., 100-4401 Still Creek Drive, Burnaby, B.C., V5C 6G9, Canada}
\author{E.~Ladizinsky}
\address{D-Wave Systems Inc., 100-4401 Still Creek Drive, Burnaby, B.C., V5C 6G9, Canada}
\author{E.~Tolkacheva}
\address{D-Wave Systems Inc., 100-4401 Still Creek Drive, Burnaby, B.C., V5C 6G9, Canada}
\author{D.~V. Averin}
\address{Department of Physics and Astronomy, University of Stony Brook, SUNY, Stony Brook, NY
11794}

\begin{abstract}
We have developed a method for extracting the high-frequency noise
spectral density of an rf-SQUID flux qubit from macroscopic resonant
tunneling (MRT) rate measurements. The extracted noise spectral
density is consistent with that of an ohmic environment up to
frequencies $\sim$ 4 GHz. We have also derived an expression for the MRT
lineshape expected for a noise spectral density consisting of such a
broadband ohmic component and an additional strongly peaked
low-frequency component. This hybrid model provides an excellent fit
to experimental data across a range of tunneling amplitudes and
temperatures.
\end{abstract}

\maketitle

Environmental noise is a critical concern in all approaches to quantum
computation, as it ultimately limits their feasibility. Understanding
the origin of noise, the way it couples to qubits, and approaches to
minimize or eliminate it will be key to any successful implementation
of a large-scale quantum computer. In this regard, experiments that
use qubits as spectrometers to probe their environment are of
particular relevance~\cite{schoelkopf2002-truncated}. In this letter we describe
an experimental approach that yields spectrosopic information
concerning the environment surrounding a superconducting flux
qubit. The method described herein is generically applicable to any
qubit whose potential energy landscape is bistable.

Many experimental techniques have been employed to characterize noise
in superconducting qubits. Rabi oscillation, spin-echo, and free
induction decay experiments all provide measurements of decoherence
times~\cite{simmonds2004-truncated,martinis2005-truncated,bennett-decoherence-2009-truncated,mcdermott-2009}.
These decoherence timescales are metrics of an aggregate response
of the qubit dynamics to the degrees of freedom of the
environment. The statistics of these environmental degrees of freedom
can be characterized by a noise spectral density $S(\omega)$. Given a
microscopic model of an environment, one can calculate $S(\omega)$
and, in turn, the decoherence timescales cited above. Inverting this
process to infer $S(\omega)$ from decoherence times can yield
model-dependent results that provide indirect probes of
$S(\omega)$. On the other hand, at very low $\omega$ one can obtain
$S(\omega)$ through direct measurements of the slow variations of
qubit
parameters~\cite{yoshihara2006-truncated,bialczak2007-truncated,lanting-lowfreq-2009-truncated,yoshihara-correlated-2010-truncated}.

Quantum tunneling experiments with strong coupling to the environment
represent an alternate means of quantifying $S(\omega)$. In this case,
$S(\omega)$ changes the observed rate of tunneling, thus providing a
direct probe of $S(\omega)$. Such dissipative tunneling has been
observed in a wide range of quantum systems, including superconducting
devices~\cite{rouse-mrt-1995},
nanomagnets~\cite{thomas-quantummagnetism-1996-truncated},
single-electron tunnel junctions with resistive
electrodes~\cite{zheng-1998}, and carbon nanotubes~\cite{bomze-2009-truncated}.
The dissipative environment of superconducting flux qubits in
particular is dominated by noise at low frequencies.  Experiments
measuring the rate of macroscopic resonant tunneling (MRT) of flux
between the two lowest energy states have consistently yielded
resonant tunneling peaks as a function of qubit bias that have a
Gaussian lineshape near their maxima. Theoretical models assuming
$S(\omega)$ is strongly peaked at small $\omega$ naturally produce
this Gaussian lineshape~\cite{amin-mrt-2008}. However, experimental
data show excess tunneling rates in the tails of the peaks that is not
explained by such theoretical
models~\cite{harris-ccjj-2010-truncated,lanting-mrt-2010-truncated,bennett-decoherence-2009-truncated}. In
this work, we quantitatively show that the non-Gaussian tails in the
MRT lineshapes can be attributed to components of $S(\omega)$ at high
frequencies. We find that $S(\omega)$, as obtained for our rf-SQUID
flux qubits, is well described by a broadband ohmic spectrum plus a
component that is strongly peaked at low $\omega$.

\begin{figure}
\includegraphics[width=3.5in]{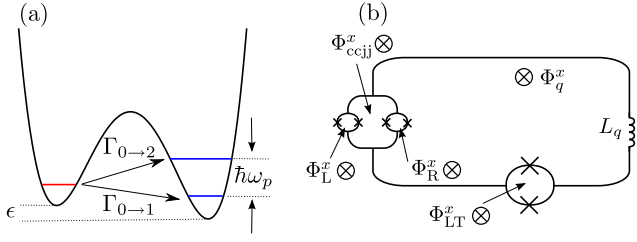}
\caption{ (a) Depiction of the double-well rf-SQUID potential and MRT
  processes discussed herein. $\Gamma_{0\rightarrow1}$ represents
  interwell tunneling between the two lowest lying
  states. $\Gamma_{0\rightarrow2}$ represents interwell tunneling
  between the lowest state in the initial well and the first excited
  state in the final well. (b) Schematic of the CCJJ rf-SQUID flux
  qubit. External fluxes applied to each closed loop indicated as
  $\Phi^x_\alpha (\alpha~\in~{L,R,\rm{ccjj},q,\rm{LT}})$.
    \label{FIG:ccjj-schematic}}
\end{figure}

In a MRT experiment~\cite{rouse-mrt-1995,harris-mrt-2008-truncated},
one prepares a flux qubit in either the right or left well
($|0\rangle$, $|1\rangle$) of the double-well flux potential of an rf
SQUID. One then measures the rate of tunneling into the opposite well
when the energy levels in the two wells are closely aligned (the
process labelled $\Gamma_{0\rightarrow1}$ in
Fig.~\ref{FIG:ccjj-schematic}(a)). For tunneling between the two lowest
energy levels, one can map the system onto a two-state Hamiltonian of
the form
\begin{equation}
H_q = -[\epsilon \sigma_z + \Delta \sigma_x]/2 - Q \sigma_z/2,
\label{EQN:canonicalhamiltonian}
\end{equation}
where $\sigma_{x,z}$ are Pauli matrices, $\Delta$ is the tunneling
amplitude, $\epsilon$ is the energy bias between the wells, and $Q$ is
a noise operator that couples the qubit to its environment. The noise
spectral density can be written as
\begin{equation}
S(\omega) =  \int dt \, e^{i\omega t}\langle Q(t)Q(0)\rangle.
\end{equation}
Hamiltonian (\ref{EQN:canonicalhamiltonian}) is valid as long as
$\epsilon,\Delta {\ll}\, \hbar \omega_p$, where $\hbar \omega_p$ is
the energy spacing to the next excited state within each well. In
writing Hamiltonian (\ref{EQN:canonicalhamiltonian}), we have
explicitly assumed that the environment couples an effective flux
signal into the qubit body.

In the limit of small tunneling amplitude, $\Delta \ll W$, where $W$
is the noise magnitude defined in Eq.~(\ref{Wep}) below, coherent
oscillations of flux between the wells are overdamped. One can then
introduce a transition rate from the left well to the right well,
$\Gamma_{0\rightarrow1}(\epsilon)\equiv \Gamma(\epsilon)$, with the
reverse rate $\Gamma_{1\rightarrow0}(\epsilon) =
\Gamma(-\epsilon)$. Under the general assumption that the noise
possesses Gaussian statistics, the tunneling rate
is given by~\cite{amin-mrt-2008,amin-mrt-2009,averin-mrt-2000}:
      \begin{equation}
      \Gamma(\epsilon) = {\Delta^2 \over 4\hbar^2}\int dt
      e^{i\epsilon t/\hbar} \exp \left\{ \int {d \omega \over 2\pi}
      S(\omega) \frac{e^{-i\omega t}{-}1}{(\hbar\omega)^2}  \right\}.
      \label{Gamma0}
      \end{equation}
If $S(\omega)$ has a dominant term that is strongly peaked at low
frequencies, which we denote $S_{LF}(\omega)$, Eq.~(\ref{Gamma0})
leads to a purely Gaussian lineshape offset by $\epsilon_p$ from the
qubit degeneracy point at $\epsilon=0$~\cite{amin-mrt-2008,harris-mrt-2008-truncated}:
\ba \Gamma(\epsilon) = \sqrt{\pi \over 8} {\Delta^2
  \over \hbar W} \exp \left\{ {-(\epsilon-\epsilon_p)^2\over 2W^2}
\right\}, \label{GaussianGamma}
\ea
where
\be
W^2 = \int {d\omega \over 2\pi} S_{LF}(\omega), \qquad
\epsilon_{p} = {\cal P} \int {d\omega \over 2\pi} {S_{LF}(\omega) \over
\hbar\omega}. \label{Wep}
\ee
(${\cal P}$ denotes principal value integration). For an environment
in thermal equilibrium at temperature $T$ and for noise strongly
peaked at low frequencies such that $\omega \ll k_BT/\hbar$ for all
relevant $\omega$ in $S_{LF}(\omega)$, the parameters $W$ and
$\epsilon_p$ are related via $\epsilon_p =
W^2/2k_BT$~\cite{amin-mrt-2008,harris-mrt-2008-truncated}. Note
that, since only cumulants of $S_{LF}(\omega)$ appear in
Eq.~(\ref{GaussianGamma}) as given by Eq.~(\ref{Wep}), MRT data cannot
provide spectroscopic information regarding $S_{LF}(\omega)$, or
indeed any feature of $S(\omega)$ on frequency scales $|\omega| < W/h$.

To account for the shape of an MRT peak away from its maximum, we
allow for both a strongly peaked low-$\omega$ component of the noise,
$S_{LF}(\omega)$, and a broadband component of the noise,
$S_{HF}(\omega)$. Writing $S(\omega) \equiv
S_{LF}(\omega){+}S_{HF}(\omega)$ and expanding the exponential
$e^{-i\omega t}$ in Eq.~(\ref{Gamma0}) up to second order in
$\omega$~\cite{amin-mrt-2008}, we obtain the following expression for
the flux tunneling rate:
\ba \Gamma(\epsilon) &=& {\Delta^2 \over 4\hbar^2}\int dt
e^{i(\epsilon-\epsilon_p) t/\hbar-W^2t^2/2\hbar^2} \nn && \exp \left\{
\int {d \omega \over 2\pi} S_{HF}(\omega) \frac{e^{-i\omega 
    t}{-}1}{(\hbar\omega)^2} \right\}.
\label{Gammae}
\ea 
Equation~(\ref{Gammae}) implies that, in contrast to $S_{LF}(\omega)$, the
shape of $S_{HF}(\omega)$ does affect the bias
dependence of the tunneling rate $\Gamma(\epsilon)$. Measurements of
$\Gamma(\epsilon)$ can therefore be used to obtain spectroscopic
information about $S_{HF}(\omega)$. To do this,
we perform two Fourier transforms. First, we define a function \be
F(t) = e^{W^2t^2/2\hbar^2} \left({4\hbar^2 \over \Delta^2}\right) \int
{d\epsilon \over 2\pi\hbar} e^{-i\epsilon t/\hbar}
\Gamma(\epsilon{+}\epsilon_p).
 \label{eqn:ft}
 \ee
Second, combining Eqs. (\ref{Gammae}) and (\ref{eqn:ft}), we express $S_{HF}(\omega)$
in terms of $F(t)$:
\be S_{HF}(\omega) =
(\hbar\omega)^2 \int dt \, e^{i\omega t} \ln F(t). \label{SHF} \ee

The environment of a flux qubit is typically in thermal equilibrium,
as shown by previous MRT experiments
\cite{harris-mrt-2008-truncated}. In this case, one can express the
noise in terms of a spectral function $J(\omega)$:
\be
S_{HF}(\omega) = \hbar^2 J (\omega)/(1 - e^{-\beta \hbar \omega}), \label{SHFeq} \ee
where $\beta \equiv 1/k_BT$ and $J(\omega)$ can be treated as an
antisymmetric function of frequency, $J(\omega) =
(S_{HF}(\omega)-S_{HF}(-\omega))/\hbar^2$. $J(\omega)$ is related to
$F(t)$ through Eq.~(\ref{SHF}):
\be J(\omega) = 2i\omega^2 \int dt \sin \omega t \ln F(t). \label{Jomega} \ee

Using Eq.~(\ref{Jomega}), one can extract $J(\omega)$ from measured
$\Gamma(\epsilon)$ versus $\epsilon$. Although $J(\omega)$ obtained in
this way does not fully vanish at low frequencies, we assume that
$S_{LF}(\omega) \gg S_{HF}(\omega)$ for $\omega < W/h$, thus ensuring that
the separation between the low- and high-frequency noise components
underlying Eq.~(\ref{Gammae}) is well defined. The MRT rate given by
Eq.~(\ref{Gammae}) is sensitive to $S_{HF}(\omega)$ in the noise
correlator only in the time interval $t< \hbar/W$. This means that the
features of $J(\omega)$ on frequency scales $\omega < W/\hbar$ cannot
be resolved in $\Gamma(\epsilon)$. In particular, $J(\omega)$ for
$\omega < W/h$ (where $S_{LF}(\omega)\neq 0$) is effectively obtained
by smooth continuation from outside this range.

An example functional form for $J(\omega)$ is that is frequently
discussed is:
\begin{equation}
J(\omega){=} \eta \omega |\omega/\omega_c|^se^{-|\omega|/\omega_c},
\label{Eqn:Jgeneral}
\end{equation}
where $\eta$ is a dimensionless parameter characterizing the strength
of the noise, $s$ describes the noise frequency dependence, and
$\omega_c$ is a high-frequency cutoff~\cite{Leggett1987-truncated}. We
assume that $\hbar\omega_c \gg k_B T, W, \epsilon$ for our work.  As
we show below, $J(\omega)$ obtained from our flux qubits is consistent
with an ohmic environment, for which $s{=}\,0$. Using
Eq.~(\ref{Eqn:Jgeneral}) with $s=0$ and substituting into
Eq.~(\ref{Gammae}) yields
 \ba
 \Gamma(\epsilon) = {\Delta_r^2 \over 4\hbar}
 \int d\tau e^{{i(\epsilon-\epsilon_p) \tau} - {W^2\tau^2/2}} \left[i
 \sinh {\tau{-}i\tau_c \over \beta/\pi} \right]^{{-\eta \over 2\pi}},\ \
  \label{GammaMixed}
\ea 
where $\Delta_r= (\pi/\beta \omega_c)^{\eta \over 4\pi} \Delta$ is the
renormalized tunneling amplitude and $\tau_c=
(\omega_c)^{-1}$. For $\eta {\ll} 4\pi$, the dependence of
$\Delta_r$ ($\simeq \Delta$) on both $T$ and $\omega_c$ is very
weak. Moreover, the role of $\tau_c$ in the integral is to remove the
divergence and its exact value does not significantly affect the
integration result. The lineshape is therefore insensitive to
$\omega_c$ for $\hbar\omega_c \gg T,W$. Equations~(\ref{GaussianGamma}) and
(\ref{GammaMixed}) are the two key predictions for the form of
$\Gamma(\epsilon)$ in the presence of $S_{LF}(\omega)$, with and
without $S_{HF}(\omega)$, respectively.


Measurements were performed with a compound-compound Josephson
junction (CCJJ) rf-SQUID flux qubit
\cite{harris-ccjj-2010-truncated}. Figure \ref{FIG:ccjj-schematic}(b)
shows a schematic of this device. Static flux biases $\Phi^x_L$ and
$\Phi^x_R$ are used to balance the critical current of the left and
right minor loops. A static flux bias $\Phi^x_{\rm{LT}}$ allows one to
adjust the inductance of the qubit.  Time dependent flux biases
$\Phi^x_q$ and $\Phi^x_{\rm{ccjj}}$ permit control of the persistent
current in the main loop $|I^p_q|$, tunneling energy $\Delta$, and
energy bias $\epsilon{=}\,2|I^p_q|(\Phi^x_q{-}\Phi^0_q)$ between the
left and right wells of the potential, where $\Phi^0_q$ is the qubit
degeneracy point. Both $\Delta$ and $|I^p_q|$ are functions of
$\Phi^x_{\rm{ccjj}}$. Time dependent external biases $\Phi^x_q(t)$ and
$\Phi^x_{\rm{ccjj}}(t)$ were delivered with high-precision room
temperature current sources through flux bias lines that were
inductively coupled to the qubit loop and the CCJJ loop, each with a
mutual inductance $\sim 2$ pH. Cold filtering on all flux bias lines
limited the available bandwidth to $\sim 5$ MHz.

The circuit was manufactured on a silicon wafer with thermal oxide,
Nb/Al/Al$_2$O$_3$/Nb trilayer junctions, and three additional Nb
wiring layers insulated from one another with planarized, high-density
plasma-enhanced chemical vapor deposited SiO$_2$. We mounted this chip
inside an Al shield on the mixing chamber stage of a dilution refrigerator
with a minimum base temperature of 21 mK. All qubit parameters were calibrated 
as described elsewhere \cite{harris-ccjj-2010-truncated}. For our qubit, we extracted 
critical current $I_c {=}\, 3.38\pm 0.01\ \mu$A, inductance $L_q {=}\, 338\pm1$ pH, 
and capacitance $C {=}\, 185 \pm 5$ fF.


\begin{figure}[t]
 \includegraphics[width=3in]{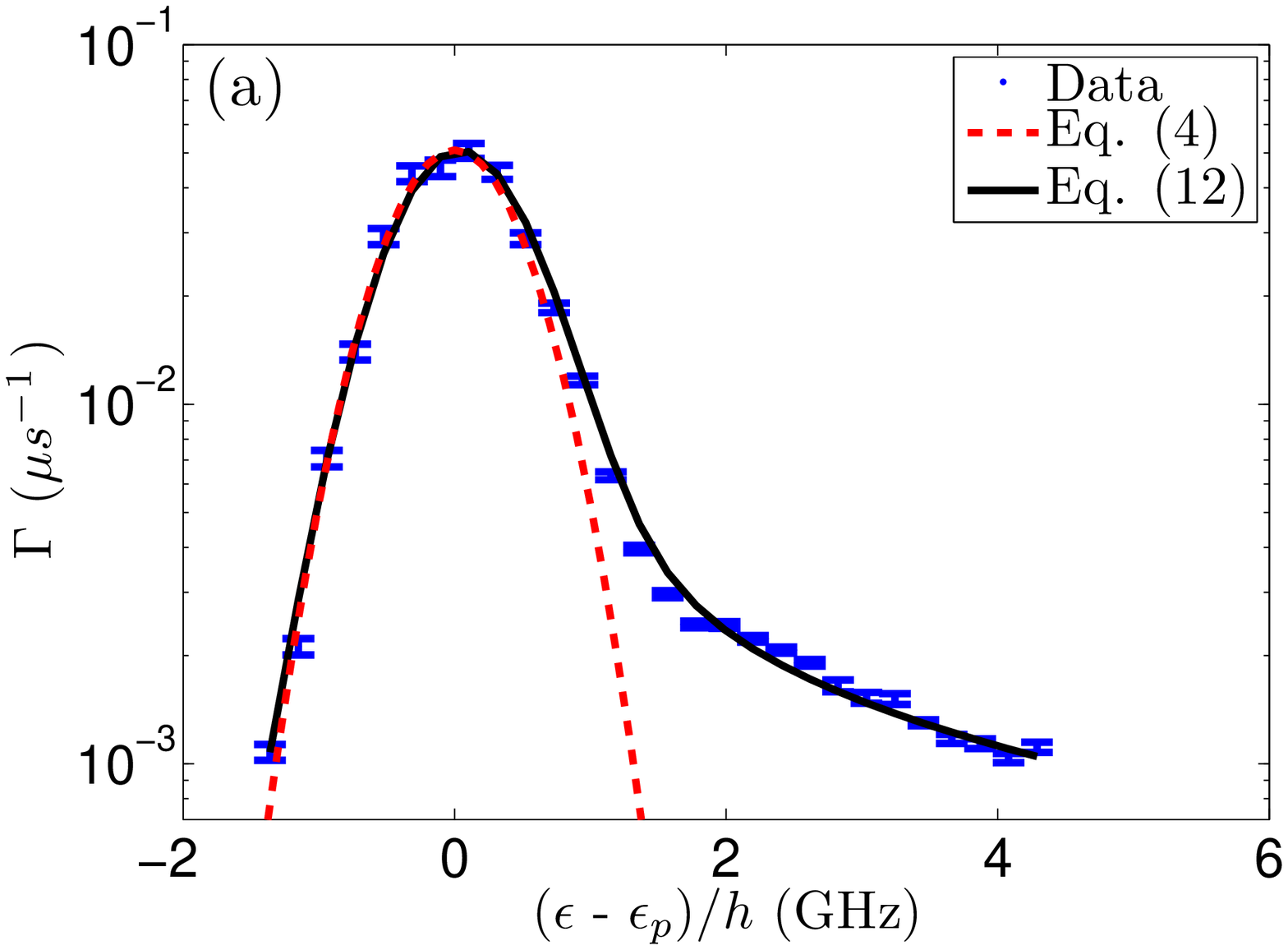}
 \includegraphics[width=3in]{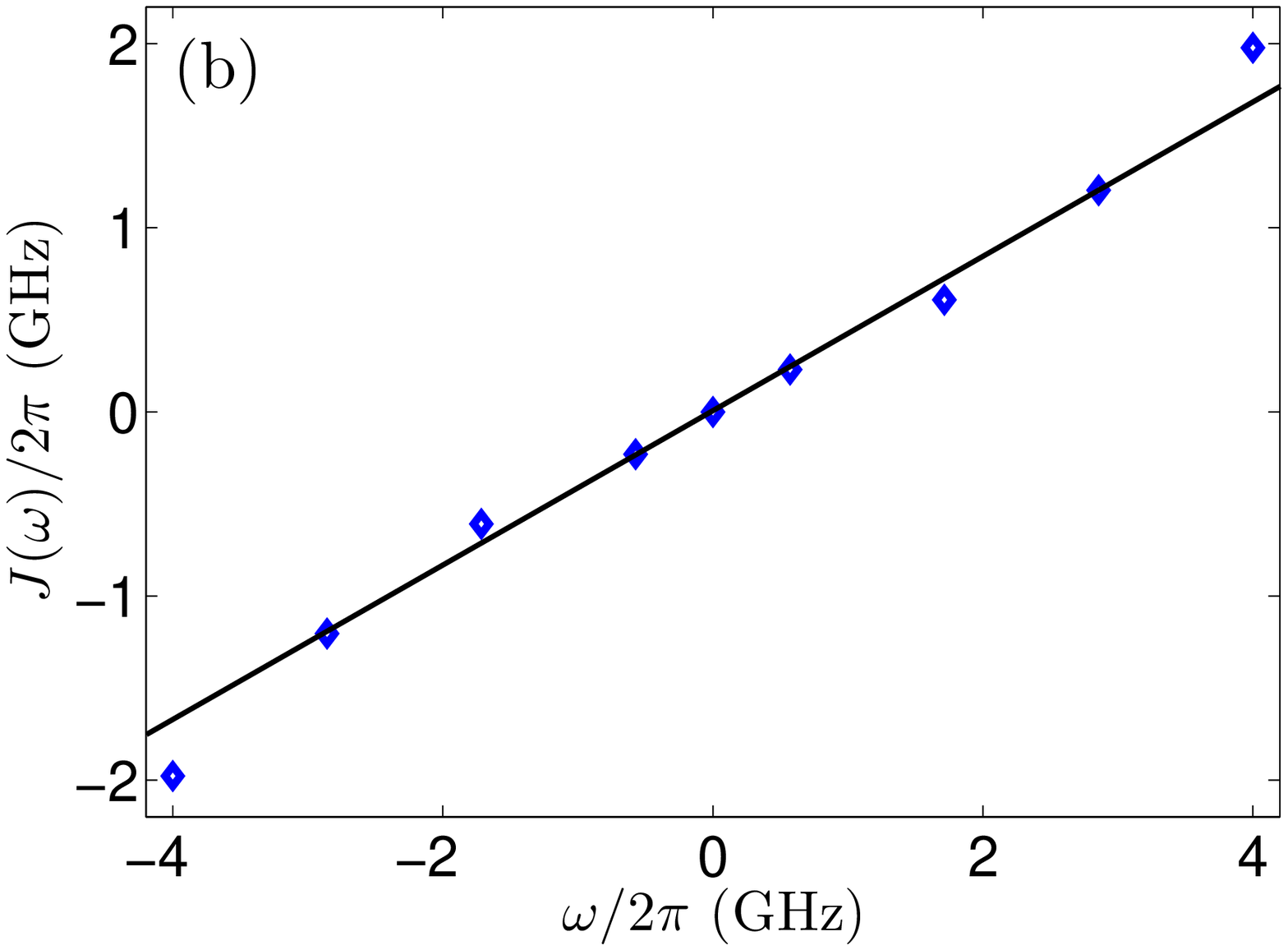}
\caption{(Color online) (a) Measured MRT rate as a function of $\epsilon$ 
at $T=21$ mK and $\Phi^x_{\rm{ccjj}}/\Phi_0 = -0.6344$.
The dashed curve is calculated using Eq.~(\ref{GaussianGamma}) with $W/h = 0.47$ 
GHz. The solid curve is calculated using Eq.~(\ref{GammaMixed}) with $W/h=0.47$ GHz 
and $\eta = 0.41$. (b) $J(\omega)$ extracted from the experimental data using 
Eq.~(\ref{Jomega}). The solid line is $J(\omega)=0.41 \omega$.}
    \label{Jw}
\end{figure}

The basic MRT rate measurement technique has been described
elsewhere~\cite{harris-mrt-2008-truncated}. Figure \ref{Jw}(a) shows example
rate measurements versus $\epsilon$. The lowest energy MRT peak is very well
separated from the next lowest resonant tunneling peak, which occurs at
$\epsilon/h \sim 13$ GHz above the first peak (see process $\Gamma_{0\rightarrow2}$ depicted in
Fig.~\ref{FIG:ccjj-schematic}(a)). For the half-decade in
$\Gamma(\epsilon)$ near the peak, the Gaussian lineshape
(\ref{GaussianGamma}) is a reasonable description of the data. To
describe the resonant peak away from its maximum, we allowed for finite
$S_{HF}(\omega)$ and extracted $J(\omega)$ from the data using
Eq.~(\ref{Jomega}). The results, plotted in Fig.~\ref{Jw}(b), agree
very well with a straight line with slope $\eta {=}\, 0.41$ up to
$|\omega/2\pi|{\approx}\, 4$ GHz, the highest frequency for which we
collected $\Gamma(\epsilon)$. Linearity of $J(\omega)$ implies an
ohmic environment. The theoretical lineshape calculated via
Eq.~(\ref{GammaMixed}) using $\eta {=}\, 0.41$ is in excellent
agreement with the experimental data [solid curve in
Fig.~\ref{Jw}(a)].

\begin{figure}[t]
 \includegraphics[width=2.5in]{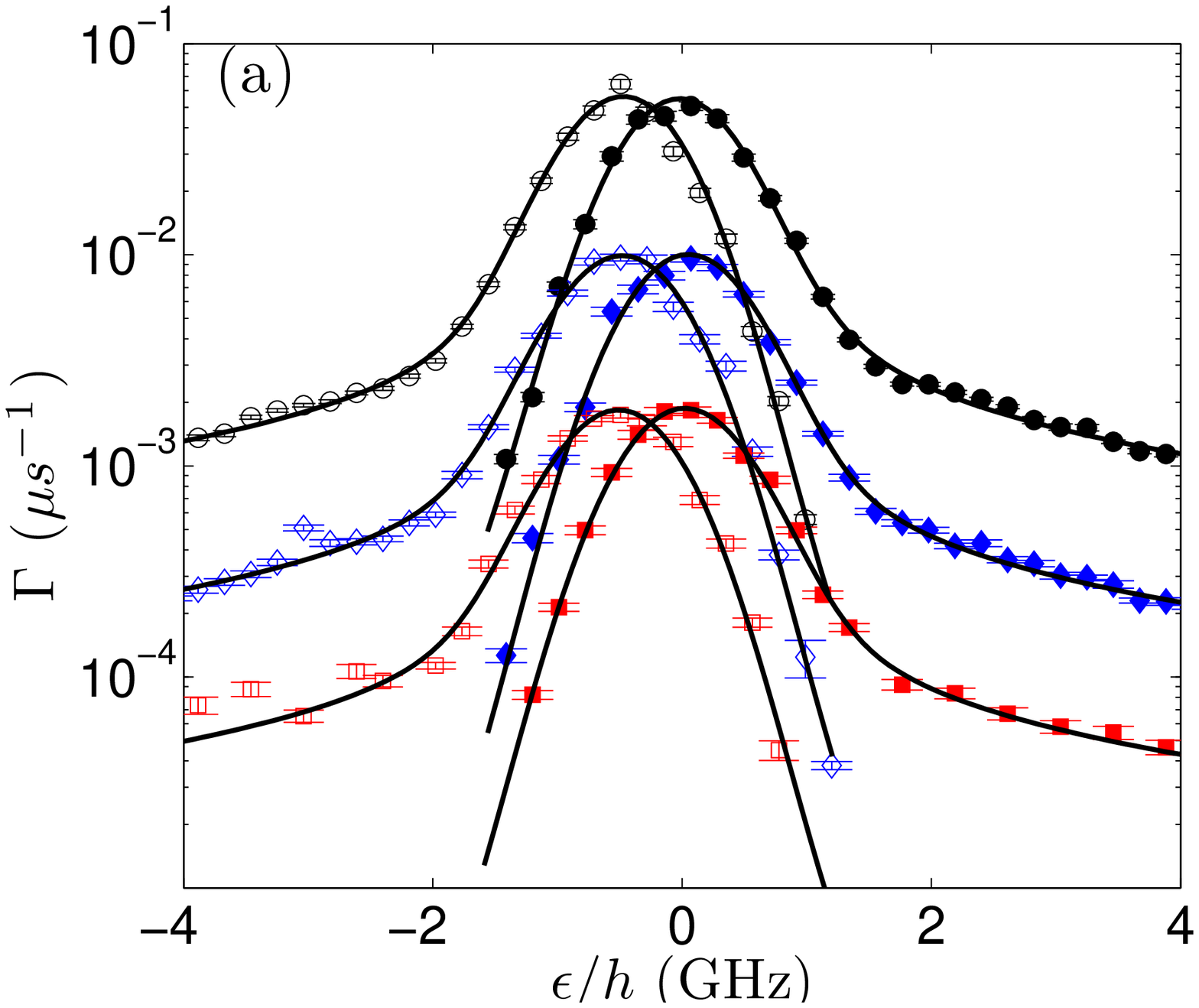}
 \includegraphics[width=2.5in]{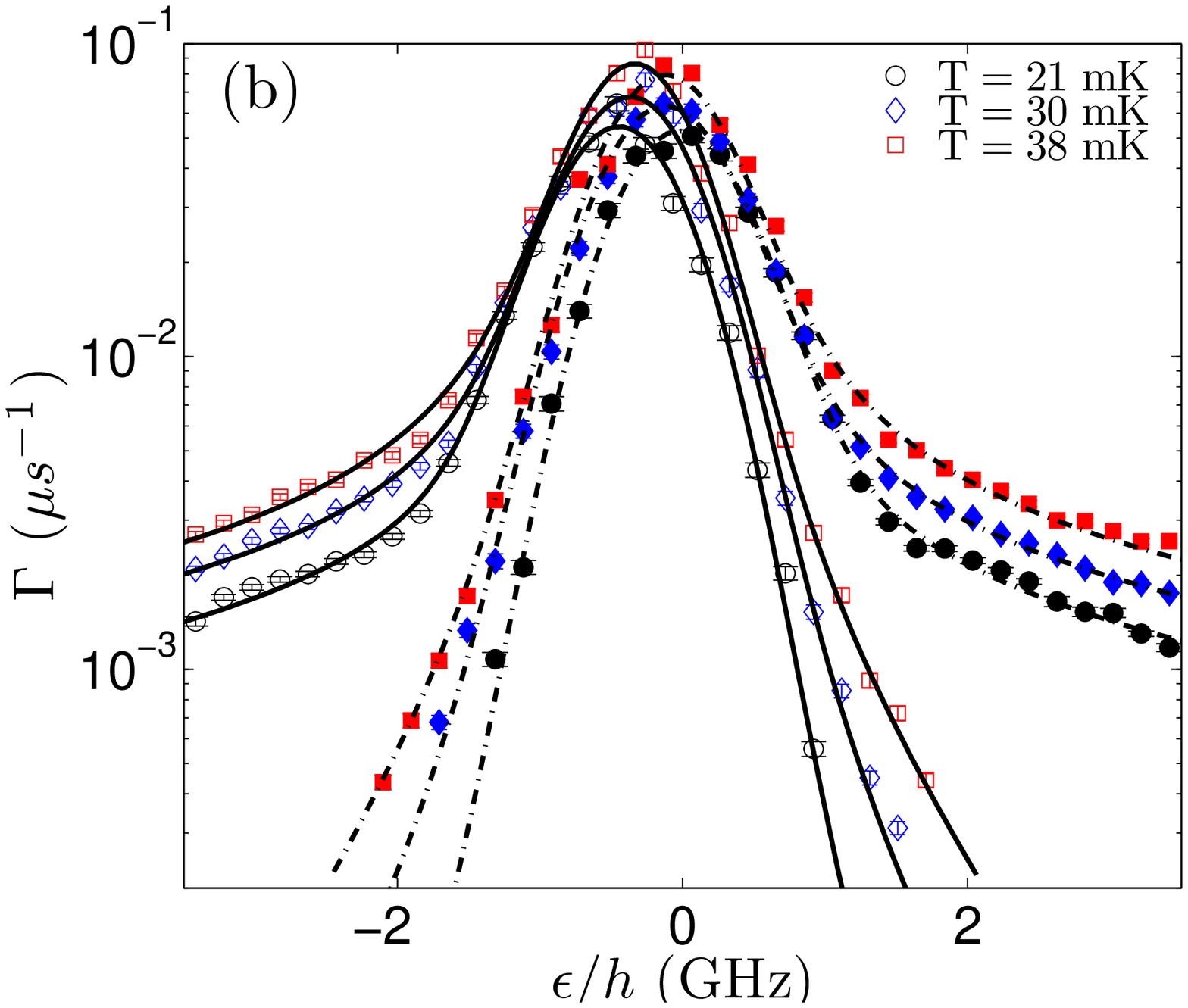}
\caption{(Color online) Example measurements of MRT rate versus
  $\epsilon$ for (a) three different barrier heights
  ($\Phi^x_{\rm{ccjj}}/\Phi_0 = -0.6344, -0.6355, -0.6365$ from top to
  bottom, respectively) at $T = 21$ mK, and (b) a single barrier
  height, $\Phi^x_{\rm{ccjj}}/\Phi_0 = -0.6344$, and three
  temperatures. The hollow (solid) symbols are $\Gamma_{0\rightarrow1}
  (\Gamma_{1\rightarrow0})$. The solid lines are a fit to
  Eq.~(\ref{GammaMixed}).}
    \label{FIG:waterfall}
\end{figure}

Given the success of an ohmic model for $S_{HF}(\omega)$, we proceeded
to directly fit a larger set of experimental data to
Eq.~(\ref{GammaMixed}).  Figure \ref{FIG:waterfall}(a) shows
measurements of $\Gamma(\epsilon)$, both $\Gamma_{0\rightarrow1}$ and
$\Gamma_{1\rightarrow0}$, for three different values of
$\Phi^x_{\rm{ccjj}}$ ($\Phi^x_{\rm{ccjj}}/\Phi_0 = -0.6344, -0.6355$
and $-0.6365$) with solid lines indicating fits to
Eq.~(\ref{GammaMixed}). We obtained fit values of $\Delta_r/h =
3.15\pm0.07, 1.38\pm0.08$ and $0.60\pm0.04$ MHz from top to bottom,
respectively. The extracted temperature was $T = 21\pm 1$ mK, in
agreement with thermometry mounted on the mixing chamber of the
dilution refrigerator. The fit width $W/h = 0.47 \pm 0.02, 0.47 \pm
0.02,$ and $0.49 \pm 0.02$ GHz from top to bottom, respectively. We
obtained $\eta = 0.41 \pm 0.03, 0.42 \pm 0.03, 0.425 \pm 0.05$ from
top to bottom, respectively.

We also performed MRT rate measurements at $\Phi^x_{\rm{ccjj}} {=}
-0.6344\ \Phi_0$ for a range of chip temperatures from 21 to 38
mK. Figure \ref{FIG:waterfall}(b) shows example measurements and fits
for three temperatures. We monitored the chip temperature via
refrigerator thermometry and confirmed it by measuring the qubit
transition width as described in~\cite{harris-ccjj-2010-truncated}.  The
temperatures extracted from the fit parameters used in Fig~\ref{FIG:waterfall}(b) match
those reported by thermometry and those obtained via qubit
transition width measurements. Fit values of $\eta$ and $W$ were relatively insensitive to chip
temperature over the range probed in our experiments. We note that the
$T$-independence of $W$ implies a $1/T$-dependence of the
linear response of the low frequency environment (typical, for
example, of a population of paramagnetic
spins)~\cite{harris-mrt-2008-truncated}. In contrast, the
$T$-independence of $\eta$ implies $T$-independence of the linear
response of the high frequency environment, consistent with an ohmic
environment.
We observed a gradual increase of $\Delta_r/h$ with $T$: $\Delta_r/h =
3.15\pm0.07, 3.45\pm0.08,$ and $ 3.6\pm0.1$ MHz for 21 mK, 30 mK and
38 mK, respectively. Only a fraction of this increase in $\Delta_r/h$
with $T$ can be accounted for by the renormalization of $\Delta$ in
Eq.~(\ref{GammaMixed}), given the best fit value of $\eta$. The rest
may be due to the influence of the higher energy levels in the flux
potential wells.




The dimensionless parameter $\eta$ characterizes the amplitude of the
high frequency noise spectral density at a given $\Phi^x_{\rm{ccjj}}$, and therefore one particular persistent current $|I^p_q|$, for our devices. 
To move closer to a physical picture of the source of high frequency noise, 
we can relate $S_{HF}(\omega)$ to an effective flux noise $S_\Phi(\omega)$ by scaling the former by the persistent current at which the measurement was performed, $S_\Phi(\omega) = S_{HF}(\omega)/(2|I^p_q|)^2$.
%
%

The source of ohmic flux noise can be parameterized as an effective
resistance $R_s$ shunting the qubit junctions. The noise spectral
density would then be
\begin{equation}
S_{\Phi}(\omega) =
\frac{2L_q^2}{R_s}{\frac{\hbar\omega}{1-e^{-\hbar\omega/k_BT}}}.
\label{EQN:ohmic-current}
\end{equation}
Using the measured qubit properties and the extracted
$\eta~\approx~0.42$, we calculate an effective shunt resistance
\begin{equation}
R_s = \frac{8|I^p_q|^2L_q^2}{\hbar\eta} \sim 20\ \rm{k}\Omega.
\end{equation}
Another potential source of ohmic noise could be external qubit flux
bias leads, each of which can be modeled as an impedance $Z_0$ coupled
to the qubit body with $M = 2$ pH. Considering the bias coupled to the
qubit body and using measured qubit properties and
$\eta~\approx~0.42$, we calculate an effective impedance
\begin{equation}
{\rm{Re}}(Z_0) = \frac{8|I^p_q|^2M^2}{\hbar\eta} \sim 1\ \Omega.
\end{equation}

Both of the ohmic sources hypothesized above predict impedances that are at
least an order of magnitude smaller than expected (independent
junction measurements suggest $R_s > 500\ \rm{k}\Omega$; we estimate
$Z_0 \sim 25\ \Omega$), making the ultimate source of the high
frequency environment uncertain.  Generally, the amplitude of the high
frequency noise should depend strongly on the details of the qubit
wiring, junction size, and strength of coupling to bias leads
depending on its source. Future measurements of this amplitude for a
variety of qubits with a range of wiring and junction sizes will allow
us to probe the ultimate source of this noise.

To summarize, we have developed and experimentally tested a method for
extracting the high frequency noise spectral density $S_{HF}(\omega)$
from MRT rate measurements on flux qubits. Our experimental data are
consistent with an ohmic spectral density up to $\omega/2\pi = 4$
GHz. We have derived a theoretical expression for the MRT lineshape
that includes both low and high frequency noise components. The
resulting model fits the experimental data very well. In particular,
this model explains tunneling rate measurements away the resonant
peak, where the model without high frequency noise fails. Our method
allows further exploration of high frequency noise in devices via its
dependence on qubit geometry and fabrication details. A systematic
study of a range of qubit designs will aid in ultimately understanding
the origin of high frequency noise in superconducting qubits.

We acknowledge fruitful discussions with F. Cioata, P. Spear,
E. Chapple, P. Chavez, C. Enderud, J. Hilton, C. Rich, G. Rose,
M. Thom, S. Uchaikin, and B. Wilson.



\end{document}